\documentclass[11pt]{article}

\usepackage{german,amsmath,amsthm,amscd,array,amssymb}

\begin{document}

\begin{center}
{\LARGE{\bf The basic cohomology
\\
\smallskip
of the twisted $N = 16$, $D = 2$ super Maxwell theory}}
\\
\bigskip\medskip
{\large{\sc B. Geyer}}$^{a}$
\footnote{Email: geyer@itp.uni-leipzig.de}
{\large{\sc and}}
{\large{\sc D. M\"ulsch}}$^{b}$
\footnote{Email: muelsch@informatik.uni-leipzig.de}
\\
\smallskip
{\it $^a$ Universit\"at Leipzig, Naturwissenschaftlich-Theoretisches Zentrum
\\
$~$ and Institut f\"ur Theoretische Physik, D--04109 Leipzig, Germany
\\
\smallskip
$\!\!\!\!\!^b$ Wissenschaftszentrum Leipzig e.V., D--04103 Leipzig, Germany}
\\
\bigskip
{\small{\bf Abstract}}
\\
\end{center}

\begin{quotation}
\noindent {\small{We consider a recently proposed two--dimensional
Abelian model for a Hodge theory, which is neither a Witten type nor a
Schwarz type topological gauge theory. It is argumented that this model
is not a good candidate for a Hodge theory because, on--shell, the BRST
Laplacian vanishes. We show, that this model allows a natural extension
such that the resulting topological theory is of Witten type and can be
identified with the twisted $N = 16$, $D = 2$ super Maxwell theory.
Furthermore, the underlying basic cohomology preserves the Hodge--type
structure and, on--shell, the BRST Laplacian does not vanish.}}
\end{quotation}

\bigskip
\begin{flushleft}
{\large{\bf 1. Introduction}}
\end{flushleft}
\medskip
It has been known for some time that two--dimensional Yang--Mills
theory without matter is exactly soluble \cite{1}. Moreover, it
has been shown that Yang--Mills theory with matter on an arbitrary
orientable two--dimensional manifold $M$ of genus $G$ and area $A$
is equivalent to a closed, orientable, string theory with target
space $M$ \cite{2}. Two--dimensional Yang--Mills theory was
revisited using a non--Abelian version of the Duistermaat--Heckman
formula in \cite{3} and a simple mapping to topological
Yang--Mills theory with an underlying $N_T = 1$ equivariantly
nilpotent shift symmetry was given. For a previous work on quantum
gauge theories in two dimensions see, e.g., \cite{4}. In this
paper we consider a Hodge type cohomological gauge theory with an
underlying $N_T = 8$ strictly nilpotent shift and co--shift
symmetry (i.e., even prior to the introduction of the gauge ghost
and anti--ghost fields), which leads to other topological
observables.

The study of such theory was motivated by a recently series of
papers \cite{5} where a class of topological gauge theories in two
dimensions was presented which is neither of Witten nor of Schwarz
type. Rather, they show some of the characteristic features of
both types of topological quantum field theories (TQFT), namely,
the form of the action turns out to be of Witten type whereas the
underlying supersymmetries are reminiscent of Schwarz type (for a
review of  TQFT, see, e.g.,~\cite{6}).

The aim of this Letter is to reveal these rather unusual properties of a TQFT.
Thereby, we focus on the simplest of the models considered in \cite{5},
namely the Euclidean Maxwell theory in two dimensions in the Feynman gauge.
Their action is given by
\begin{equation}
\label{1.1}
S = \int_E d^2x\, \Bigr\{
\hbox{$\frac{1}{4}$} F^{ab}(A) F_{ab}(A) +
\hbox{$\frac{1}{2}$} \partial^a A_a \partial^b A_b +
\partial^a \bar{C} \partial_a C \Bigr\}\,,
\end{equation}
with $F_{ab}(A) = \partial_a A_b - \partial_b A_a$, where
$A_a$ is the Abelian gauge field and $C$, $\bar{C}$ are the gauge
(anti)ghost fields, respectively.

In \cite{4} it has been shown that the action (\ref{1.1}) is not
only invariant under the BRST symmetry, generated by $\Omega$, but
also under a co--BRST symmetry, generated by $^\star \Omega$,
which, together with the BRST Laplacian $W$, obey the following
BRST--complex:
\begin{equation}
\label{1.2}
\Omega^2 = 0,
\qquad
^\star \Omega^2 = 0,
\qquad
W = \{ \Omega, \,^\star \Omega \} \neq 0,
\qquad
[ \Omega, W ] = 0,
\qquad
[ \,^\star \Omega, W ] = 0.
\end{equation}

Representations of this superalgebra for the first time have been
considered in \cite{7}. Since $\Omega$ and $^\star \Omega$ are
nilpotent hermitian operators they are realized in a Krein space
$\cal K$ \cite{8} whose non--degenerate indefinite scalar product
$\langle \chi | \psi \rangle := (\chi, J \psi)$ is defined by the
help of a self--adjoint metric operator $J \neq 1$, $J^2 = 1$.
With respect to the inner product $(~,~)$ the operators
$\Omega$ and $^\star \Omega = \pm J \Omega J$ are adjoint to each
other, $(\chi, \,^\star \Omega \psi) = (\Omega \chi, \psi)$, but
with respect to the scalar product $\langle \,\, | \; \rangle$
they are self--adjoint.
From these definitions one obtains a remarkable correspondence
between the BRST cohomology and the de Rham cohomology \cite{9}:
\begin{alignat*}{4}
&\hbox{BRST operator}
&\quad&
\Omega,
&\qquad\qquad&
\hbox{differential}
&\quad&
d,
\\
&\hbox{co--BRST operator}
&\quad&
^\star\Omega = \pm J \Omega J,
&\qquad\qquad&
\hbox{co--differential}
&\quad&
\delta = \pm \star d \star,
\\
&\hbox{duality operation}
&\quad&
J,
&\qquad\qquad&
\hbox{Hodge star}
&\quad&
\star,
\\
&\hbox{BRST Laplacian}
&\quad&
W = \{ \Omega, \,^\star\Omega \},
&\qquad\qquad&
\hbox{Laplacian}
&\quad&
\Delta = \{ d, \delta \}.
\end{alignat*}

 Hence, the action (\ref{1.1}) provides a Hodge type field
theoretical model in two dimensions. However, owing to the absence
of a shift and co--shift symmetry, the action is not of Witten
type and, because the Maxwell action in a gravitational background
is not metric independent, it is also not of Schwarz type.
Differently, the topological nature of that model is a consequence
of the fact that in two dimensions there are no propagating
degrees of freedom associated with the gauge field. On the other
hand, as there is no topological supersymmetry, on--shell, the BRST Laplacian
vanishes. This is, in fact, an unsatisfactory property of a Hodge type
theory because their physical states should lie entirely in the set of
harmonic states, i.e., the set of the zero modes of the BRST Laplacian.
Therefore, in order to incorporate a topological supersymmetry into
that model, it will be shown that it can be regarded as part of a more
complex topological model of Witten type, namely the twisted $N = 16$,
$D = 2$ super Maxwell theory with global symmetry group $SU(4)$, whose
{\it basic cohomology} \cite{10} possesses actually a Hodge type structure.

The Letter is organized as follows: In Sect. 2, as a first step,
we substitute in (\ref{1.1}) the Maxwell action by the
cohomological action of twisted $N = 16$, $D = 2$ super Maxwell
theory with global symmetry group $SU(4)$. And we show that the
BRST complex of the $8$ twisted scalar supercharges, i.e., the
generators of the shift and co--shift symmetries, is really of
Hodge type. In Sect. 3, as a second step, we complete the
cohomological action by introducing the ordinary gauge fixing
terms and verify that the basic cohomology, i.e., the BRST complex
including also the ordinary gauge symmetry, preserves the
underlying Hodge type structure. In Sect. 4 we construct topological
observables for that theory.

\bigskip
\begin{flushleft}
{\large{\bf 2. The BRST complex of the twisted $N = 16$, $D = 2$
super Maxwell theory with global symmetry group $SU(4)$}}
\end{flushleft}
\medskip
As mentioned above, in order to avoid the vanishing of the BRST Laplacian
of the topological model proposed in \cite{5} we view the Maxwell action
in (\ref{1.1}) as the classical part of the twisted action of $N = 16$,
$D = 2$ super Maxwell theory with global symmetry group $SU(4)$
(see, e.g., \cite{11} where the non--Abelian extension of that action has
been constructed),
\begin{align}
\label{2.1}
S_T^{(N_T = 8)} = \int_E d^2x\, \Bigr\{&
\hbox{$\frac{1}{4}$} F^{ab}( A + i V ) F_{ab}( A - i V ) +
\hbox{$\frac{1}{2}$} \partial^a V_a \partial^b V_b
\nonumber
\\
& + \hbox{$\frac{1}{8}$} \partial^a M_{\alpha\beta}
\partial_a M^{\alpha\beta} -
\epsilon^{ab} \zeta_\alpha \partial_a \psi_b^\alpha -
\eta_\alpha \partial^a \psi_b^\alpha \Bigr\}.
\end{align}
Here, $V_a$ is a co--vector field which is combined with $A_a$ to
forms the complexified gauge fields $A_a \pm i V_a$, i.e., the
action (\ref{2.1}) localizes onto the moduli space of complexified
flat connections. Furthermore, we have introduced a
$SU(4)$--quartet of Grassmann--odd vector fields
$\psi_a^\alpha$, two $SU(4)$--quartets of Grassmann--odd scalar
fields, $\eta_\alpha$ and $\zeta_\alpha$, which
transform as the fundamental and its complex conjugate
representation of $SU(4)$, respectively, and a $SU(4)$--sextet of
Grassmann--even complex scalar fields $M_{\alpha\beta} =
\hbox{$\frac{1}{2}$} \epsilon_{\alpha\beta\gamma\delta}
M^{\gamma\delta}$, which transform as the second--rank complex
selfdual representation of $SU(4)$, where $\alpha = 1,2,3,4$
denotes the internal group index of $SU(4)$. $\epsilon_{ab}$
is the antisymmetric Levi--Civita tensor in $D = 2$.

Let us notice that the action (\ref{2.1}) can be obtained from the
Euclidean $N = 16$, $D = 2$ super Maxwell theory with R--symmetry
group $SU(4) \otimes U(1)$ by twisting the Euclidean rotation
group $SO_E(2) \sim U_E(1)$ in $D = 2$ by the $U(1)$ of the
R--symmetry group (by simply putting together both $U(1)$
charges), thereby leaving the group $SU(4)$ intact \cite{12}.

The action (\ref{2.1}) is invariant under the following discrete
Hodge type $\star$--symmetry, defined by the replacements
\begin{equation}
\label{2.2}
\varphi \equiv \begin{bmatrix}
\partial_a & A_a & V_a &
\\
\psi_a^\alpha & \eta_\alpha & \zeta_\alpha & M^{\alpha\beta}
\end{bmatrix}
\quad \Rightarrow \quad
\star \varphi = \begin{bmatrix}
\epsilon_{ab} \partial^b & \epsilon_{ab} A^b &
- \epsilon_{ab} V^b &
\\
- i \psi_a^\alpha & - i \zeta_\alpha & i \eta_\alpha &
- M^{\alpha\beta}
\end{bmatrix},
\end{equation}
with the property $\star (\star \varphi) = \pm \varphi$.

Let us now describe the full set of twisted supersymmetry transformations
which leave the action (\ref{2.1}) invariant. The transformation rules for
the {\it on--shell} shift symmetries $Q^\alpha$ are
\begin{align}
\label{2.3}
&Q^\alpha A_a = \psi_a^\alpha,
\nonumber
\\
&Q^\alpha V_a = - i \psi_a^\alpha,
\nonumber
\\
&Q^\alpha M_{\beta\gamma} = 2 i \delta^\alpha_{~[\beta}
\zeta_{\gamma]},
\nonumber
\\
&Q^\alpha \psi_a^\beta = - i \epsilon_{ab} \partial^b M^{\alpha\beta},
\nonumber
\\
&Q^\alpha \eta_\beta = - i \delta^\alpha_{~\beta}
\partial^a V_a,
\nonumber
\\
&Q^\alpha \zeta_\beta = \hbox{$\frac{1}{2}$}
\delta^\alpha_{~\beta} \epsilon^{ab} F_{ab}(A - i V).
\qquad\qquad\qquad\quad\qquad\qquad\qquad
\end{align}
From combining $Q^\alpha$ with the above displayed Hodge type
$\star$--symmetry one gets the corresponding transformation rules
for the {\it on--shell} co--shift symmetries $^\star Q^\alpha$,
i.e., $^\star Q^\alpha \varphi = \pm \star Q^\alpha \star
\varphi$, where the signs are the same as in the relation $\star
(\star \varphi) = \pm \varphi$.

Furthermore, by making use of the identity $\epsilon_{ab}
\partial_c + \epsilon_{bc} \partial_a + \epsilon_{ca} \partial_b = 0$,
one simply verifies that (\ref{2.1}) is also invariant under the following
{\it on--shell} vector supersymmetries $Q_{a \alpha}$,
\begin{align}
\label{2.4}
&Q_{a \alpha} A_b = \delta_{ab} \eta_\alpha -
\epsilon_{ab} \zeta_\alpha,
\nonumber
\\
&Q_{a \alpha} V_b = - i \delta_{ab} \eta_\alpha -
i \epsilon_{ab} \zeta_\alpha,
\nonumber
\\
&Q_{a \alpha} M^{\beta\gamma} = 2 i \epsilon_{ab}
\delta_\alpha^{~[\beta} \psi^{b \gamma]},
\nonumber
\\
&Q_{a \alpha} \psi_b^\beta = i \delta_\alpha^{~\beta}
\delta_{ab} \partial^c V_c -
\delta_\alpha^{~\beta} \partial_a ( A_b + i V_b ) +
\delta_\alpha^{~\beta} \partial_b ( A_a - i V_a ),
\nonumber
\\
&Q_{a \alpha} \eta_\beta = i \epsilon_{ab}
\partial^b M_{\alpha\beta},
\nonumber
\\
&Q_{a \alpha} \zeta_\beta = i \partial_a M_{\alpha\beta}.
\end{align}
In principle, owing to the Hodge type $\star$--symmetry, one can
also introduce {\it on--shell} co--vector supersymmetries $^\star
Q_{a \alpha}$, namely, similar as before, according to
$^\star Q_{a \alpha} \varphi = \pm \star Q_{a \alpha}
\star \varphi$. However, they become $i$ times the {\it on--shell} vector
supersymmetries $Q_{a \alpha}$, i.e.,
$^\star Q_{a \alpha} = i Q_{a \alpha}$. Hence, it holds
$( Q^\alpha, \,^\star Q^\alpha, Q_{a \alpha} ) S_T^{(N_T = 8)} = 0$,
and the total number of (real) supercharges is actually $N = 16$.

By an explicit calculation one establishes that the $8$ scalar
supercharges $Q^\alpha$ and $^\star Q^\alpha$, being interrelated
by the $\star$--operation, together with the $8$ vector supercharges
$Q_{\mu \alpha}$, obey the following topological superalgebra {\it on--shell},
\begin{equation*}
\{ Q^\alpha, Q^\beta \} \doteq 0,
\qquad
\{ Q^\alpha, \,^\star Q^\beta \} \doteq - 2 \delta_G(M^{\alpha\beta}),
\qquad
\{ \,^\star Q^\alpha, \,^\star Q^\beta \} \doteq 0,
\end{equation*}
\begin{equation*}
\{ Q^\alpha, Q_{a \beta} \} \doteq
- 2 \delta^\alpha_{~\beta} (
\partial_a - \delta_G(A_a - i V_a) ),
\qquad
\{ \,^\star Q^\alpha, Q_{a \beta} \} \doteq
- 2 \delta^\alpha_{~\beta} (
\partial_a - \delta_G(A_a + i V_a) ),
\end{equation*}
where the field--dependent gauge transformations $\delta_G(\varphi)$,
with $\varphi = (M^{\alpha\beta}, A_a \pm i V_a)$, are defined by
$\delta_G(\varphi) A_a = \partial_a \varphi$, and zero otherwise.
(The symbol $\doteq$ means that the corresponding relation is satisfied
only {\it on--shell}.)

Finally, let us show that the twisted action (\ref{2.1}) is
actually of Witten type. For that purpose we break down the global
symmetry group $SU(4)$ in such way that the resulting action
splits into a topological term ($Q$--cocyle) and a $Q$--exact
term, where $Q$ is a certain linear combination of $Q^\alpha$ and
$^\star Q^\alpha$, and that it possesses a discrete Hodge type
$\star$--symmetry. This is precisely what we need in order to put
the theory on a two--dimensional compact Riemannian manifold.

To begin with, let us consider the following $N_T = 1$ topological
action in five dimensions,
\begin{align}
\label{2.5}
S_T^{(N_T = 1)} = \int_E d^5x\, \Bigr\{&
\hbox{$\frac{1}{4}$} F^{AB}( A + i V ) F_{AB}( A - i V ) +
\hbox{$\frac{1}{2}$} \partial^A V_A \partial^B V_B
\nonumber
\\
& - \hbox{$\frac{1}{8}$} i \epsilon_{ABCDE} \chi^{AB}
\partial^C \chi^{DE} -
i \chi^{AB} \partial_A \psi_B - i \eta \partial^A \psi_A \Bigr\},
\end{align}
which is build up from the Abelian gauge field $A_A$, the co--vector field
$V_A$ and the Grassmann--odd scalar, vector and antisymmetric tensor fields
$\eta$, $\psi_A$ and $\chi_{AB}$, respectively. Here, the space
index $A$ runs from $1$ to $5$ and $\epsilon_{ABCDE}$ is the complete
antisymmetric Levi--Civita tensor in $D = 5$.
This action can be obtained from the Euclidean $N = 2$, $D = 5$ super Maxwell
theory with R--symmetry group $SO(5)$ by twisting the Euclidean rotation group
$SO_E(5)$ in $D = 5$ by the R--symmetry group (for details, we refer to
\cite{11}, where the non--Abelian extension of that action was given).

The full set of twisted {\it on--shell} supersymmetry transformations,
generated by the scalar, vector and antisymmetric tensor supercharges
$Q$, $Q_A$ and $Q_{AB}$, respectively, are given by
\begin{align}
&Q A_A = \psi_A,
\nonumber
\\
&Q V_A = - i \psi_A,
\nonumber
\\
&Q \eta = - \partial^A V_A,
\nonumber
\\
&Q \psi_A = 0,
\nonumber
\\
&Q \chi_{AB} = - i F_{AB}(A - i V),
\label{2.6}
\\
\nonumber
\\
&Q_A A_B = \delta_{AB} \eta - \chi_{AB},
\nonumber
\\
&Q_A V_B = - i \delta_{AB} \eta - i \chi_{AB},
\nonumber
\\
&Q_A \eta = 0,
\nonumber
\\
&Q_A \psi_B = \delta_{AB} \partial^C V_C +
i \partial_A ( A_B + i V_B ) - i \partial_B ( A_A - i V_A ),
\nonumber
\\
&Q_A \chi_{BC} = - \hbox{$\frac{1}{2}$} \epsilon_{ABCDE} F^{DE}(A + i V)
\label{2.7}
\\
\intertext{and}
&Q_{AB} A_C = - \delta_{C[A} \psi_{B]} -
\hbox{$\frac{1}{2}$} \epsilon_{ABCDE} \chi^{DE},
\nonumber
\\
&Q_{AB} V_C = - i \delta_{C[A} \psi_{B]} +
\hbox{$\frac{1}{2}$} i \epsilon_{ABCDE} \chi^{DE},
\nonumber
\\
&Q_{AB} \eta = - i F_{AB}(A + i V),
\nonumber
\\
&Q_{AB} \psi_C = - \hbox{$\frac{1}{2}$} i \epsilon_{ABCDE} F^{DE}(A - i V),
\nonumber
\\
&Q_{AB} \chi_{CD} = \delta_{C[A} \delta_{B]D} \partial^E V_E -
i \delta_{[C[A} ( \partial_{B]} ( A_{D]} - i V_{D]} ) -
\partial_{D]} ( A_{B]} + i V_{B]} ) ).
\label{2.8}
\end{align}
By the help of the identity $\frac{1}{2} \epsilon_{ABCGH} \epsilon^{DEFGH} =
\delta_{[A}^D \delta_B^E \delta_{C]}^F$ it is straightforward, but tedious
to prove that the action (\ref{2.5}) is really left invariant under the
above transformations.

Let us now group the components of the $5$--dimensional fields
$A_A$, $V_A$, $\chi^{AB}$, $\psi_A$, $\eta$ into those of the
$2$--dimensional fields $A_a$, $V_a$, $M_{\alpha\beta}$,
$\eta_\alpha$, $\zeta_\alpha$, $\psi_a^\alpha$ according to
\begin{alignat*}{4}
&M_{12} = A_3 + i V_3,
&\qquad
&M_{31} = A_4 + i V_4,
&\qquad
&M_{14} = A_5 + i V_5,
&
\\
&M_{34} = A_3 - i V_3,
&\qquad
&M_{42} = A_4 - i V_4,
&\qquad
&M_{23} = A_5 - i V_5,
&
\\
&\eta_1 = i \eta,
&\qquad
&\eta_2 = i \chi^{45},
&\qquad
&\eta_3 = - i \chi^{53},
&\qquad
&\eta_4 = i \chi^{34},
\\
&\zeta_1 = \hbox{$\frac{1}{2}$} i \epsilon_{ab} \chi^{ab},
&\qquad
&\zeta_2 = - i \psi_3,
&\qquad
&\zeta_3 = i \psi_4,
&\qquad
&\zeta_4 = - i \psi_5,
\\
&\psi_a^1 = \psi_a,
&\qquad
&\psi_a^2 = \epsilon_{ab} \chi^{b 3},
&\qquad
&\psi_a^3 = - \epsilon_{ab} \chi^{b 4},
&\qquad
&\psi_a^4 = \epsilon_{ab} \chi^{b 5},
\end{alignat*}
where $(A_A, V_A)$ agrees with $(A_a, V_b)$ for $A = 1,2$. Then,
it is easily seen that by dimensional reducing the action
(\ref{2.5}) and the supersymmetry transformations
(\ref{2.6})--(\ref{2.8}) onto two dimensions, one arrives
precisely at the $N_T = 8$ topological action (\ref{2.1}) and the
supersymmetry transformations (\ref{2.3}) and (\ref{2.4}).

On the other hand, on--shell, upon using the equation of motion
for $\eta$, the action (\ref{2.5}) can be written as a sum of a
topological term ($Q$--cocycle) and a $Q$--exact term,
\begin{equation}
\label{2.9}
S_T^{(N_T = 1)} \doteq - \int_E d^5x\, \Bigr\{
\hbox{$\frac{1}{8}$} i \epsilon_{ABCDE}
\chi^{AB} \partial^C \chi^{DE} \Bigr\} + Q \Psi,
\qquad
Q^2 \doteq 0,
\end{equation}
with the gauge fermion
\begin{equation*}
\Psi = \int_E d^5x\, \Bigr\{
\hbox{$\frac{1}{4}$} i \chi^{AB} F_{AB}( A + i V ) -
\hbox{$\frac{1}{2}$} \eta \partial^A V_A \Bigr\}.
\end{equation*}
Hence, by reducing also (\ref{2.9}) in the same way as above into
two dimensions we get exactly the decomposition of the action
(\ref{2.1}) we are looking for, consisting of a topological term
($Q$--cocycle) and of a $Q$--exact term.

More precisely, if we group the components
$A_M$, $V_M$, $\psi_M$, $\chi_{aM}$, $\chi_{MN}$,
($M,N = 3,4,5$) into $SU(2)$ triplets
$M^{ij}$, $N^{ij}$, $\rho^{ij}$, $\chi_a^{~ij}$, $\lambda^{ij}$ ($i,j = 1,2$)
and identify $\frac{1}{2} \epsilon^{ab} \chi_{ab}$ with the $SU(2)$
singlet $\zeta$, we obtain the following $N_T = 8$ topological action
with a residual global symmetry group $SU(2) \otimes U(1)$,
\begin{align}
\label{2.10}
S_T^{(N_T = 8)} = \int_E d^2x\, \Bigr\{&
\hbox{$\frac{1}{4}$} F^{ab}( A + i V ) F_{ab}( A - i V ) +
\hbox{$\frac{1}{2}$} \partial^a V_a \partial^b V_b
\nonumber
\\
& + \hbox{$\frac{1}{2}$} \partial^a ( M_{ij} + i N_{ij} )
\partial_a ( M^{ij} - i N^{ij} )
\nonumber
\\
& - i \epsilon^{ab} \lambda_{ij} \partial_a \chi_b^{~ij} -
i \rho_{ij} \partial^a \chi_a^{~ij} -
i \epsilon^{ab} \zeta \partial_a \psi_b -
i \eta \partial^a \psi_a \Bigr\}.
\end{align}
The transformation rules for the {\it on--shell} shift symmetry $Q$ are
given by
\begin{alignat*}{2}
&Q A_a = \psi_a,
&\qquad
&Q V_a = - i \psi_a,
\\
&Q M^{ij} = \rho^{ij},
&\qquad
&Q N^{ij} = - i \rho^{ij},
\\
&Q \lambda^{ij} = 0,
&\qquad
&Q \rho^{ij} = 0,
\\
&Q \eta = - \partial^a V_a,
&\qquad
&Q \zeta = - \hbox{$\frac{1}{2}$} i \epsilon^{ab} F_{ab}(A - i V),
\\
&Q \psi_a = 0,
&\qquad
&Q \chi_a^{~ij} = - i \partial_a ( M^{ij} - i N^{ij} ).
\end{alignat*}
Moreover, the action (\ref{2.10}) is also invariant under the following
duality $\star$--operation, which maps $Q$ to $^\star Q$,
\begin{equation*}
\label{2.12}
\varphi \equiv \begin{bmatrix}
\partial_a & A_a & V_a
\\
M^{ij} & N^{ij} &
\\
\psi_a & \eta & \zeta
\\
\chi_a^{~ij} & \rho^{ij} & \lambda^{ij}
\end{bmatrix}
\quad \Rightarrow \quad
\star \varphi = \begin{bmatrix}
\epsilon_{ab} \partial^b & \epsilon_{ab} A^b & - \epsilon_{ab} V^b
\\
- M^{ij} & - N^{ij} &
\\
- i \psi_a & - i \zeta & i \eta
\\
- i \chi_a^{~ij} & - i \lambda^{ij} & i \rho^{ij}
\end{bmatrix}.
\end{equation*}
Hence, as anticipated, after breaking down the $SU(4)$--symmetry
to $SU(2) \otimes U(1)$ the Hodge type structure of the theory is
still preserved.
Obviously, because the maximal number of scalar supercharges of twisted
$N = 16$, $D = 2$ super Maxwell theory is $N_T = 8$, the generator $Q$ must
be a certain linear combination of $Q^\alpha$ and $^\star Q^\alpha$.

Up to now we have shown that the BRST complex of the twisted $N = 16$,
$D = 2$ super Maxwell theory possesses actually a Hodge type structure.
Moreover, we have verified that the cohomological action is of Witten type
and that it corresponds to the case with the maximal number $N_T = 8$ of
scalar supercharges and the largest possible global symmetry group $SU(4)$.
On the other hand, we could also perform a topological twist of $N = 8$,
$D = 2$ super Maxwell theory, whose underlying BRST complex possesses a
Hodge type structure, too. In that case we would get the minimal
number of $N_T = 4$ scalar supercharges and the lowest possible
global symmetry group $SU(2)$ \cite{13}.

\bigskip
\begin{flushleft}
{\large{\bf 3. The Hodge type structure of the basic cohomology}}
\end{flushleft}
\medskip
The twisted action (\ref{2.1}) has still an ordinary gauge symmetry,
which we have not considered yet. Now, we want to show that by adding
to (\ref{2.1}) the usual gauge fixing and (anti)ghost dependent terms the
Hodge type structure of the underlying {\it basic cohomology}, i.e., the
BRST complex including also the ordinary gauge symmetry, is preserved.
To this end we introduce two $SU(4)$--quartets of Grassmann--odd
scalar ghost and antighost fields, $C^\alpha$ and $\bar{C}_\alpha$, which
transform as the fundamental and its complex conjugate representation of
$SU(4)$, respectively.

Then, the complete gauge--fixed action reads
\begin{align}
\label{3.1}
S^{(N_T = 8)} = \int_E d^2x\, \Bigr\{&
\hbox{$\frac{1}{4}$} F^{ab}(A + i V) F_{ab}(A - i V) +
\hbox{$\frac{1}{2}$} \partial^a (A_a + i V_a) \partial^b (A_b - i V_b)
\nonumber
\\
& + \hbox{$\frac{1}{8}$} \partial^a M_{\alpha\beta}
\partial_a M^{\alpha\beta} -
\epsilon^{ab} \zeta_\alpha \partial_a \psi_b^\alpha -
\eta_\alpha \partial^a \psi_b^\alpha +
\partial^a \bar{C}_\alpha \partial_a C^\alpha \Bigr\}.
\end{align}
This action, in spite of the fact that the gauge symmetry is fixed, exhibits
still an invariance under the following bosonic symmetry,
\begin{align}
\label{3.2}
&W^{\alpha\beta} A_a = - 2 \partial_a M^{\alpha\beta},
\nonumber
\\
&W^{\alpha\beta} M_{\gamma\delta} = - 4 \delta^\alpha_{~[\gamma}
\delta^\beta_{~\delta]} \partial^a A_a,
\nonumber
\\
&W^{\alpha\beta} \psi_a^\gamma = 2 i \epsilon^{\alpha\beta\gamma\delta}
( \partial_a \zeta_\delta +
\epsilon_{ab} \partial^b \eta_\delta ),
\end{align}
where we have written down only the non--vanishing transformation
rules. However, below it will be shown that the generator
$W^{\alpha\beta}$ of that symmetry is just the BRST Laplacian.

Furthermore, in order to ensure the nilpotency of the BRST and
co--BRST operators, we introduce a set of auxiliary fields,
namely, the $SU(4)$--singlets of Grassmann--even scalar fields
$B$, $\bar{B}$ and $G$, $\bar{G}$. By the help of these additional
fields the action (\ref{3.1}) can be rewritten as
\begin{align}
\label{3.3}
S^{(N_T = 8)} = \int_E d^2x\, \Bigr\{&
\hbox{$\frac{1}{4}$} i \epsilon^{ab} B F_{ab}(A + i V) -
\hbox{$\frac{1}{4}$} i \epsilon^{ab} \bar{B} F_{ab}(A - i V) -
\hbox{$\frac{1}{2}$} \bar{B} B
\nonumber
\\
& + \hbox{$\frac{1}{2}$} i G \partial^a (A_a + iV_a ) -
\hbox{$\frac{1}{2}$} i \bar{G} \partial^a (A_a - i V_a) -
\hbox{$\frac{1}{2}$} \bar{G} G
\phantom{\frac{1}{2}}
\nonumber
\\
& + \hbox{$\frac{1}{8}$} \partial^a M_{\alpha\beta}
\partial_a M^{\alpha\beta} -
\epsilon^{ab} \zeta_\alpha \partial_a \psi_b^\alpha -
\eta_\alpha \partial^a \psi_b^\alpha +
\partial^a \bar{C}_\alpha \partial_a C^\alpha \Bigr\},
\end{align}
and the Hodge type $\star$--symmetry (\ref{2.2}) must be supplemented by
the following replacements,
\begin{equation}
\label{3.4}
\varphi \equiv \begin{bmatrix}
B & C^\alpha & G
\\
\bar{B} & \bar{C}_\alpha & \bar{G}
\\
\end{bmatrix}
\quad \Rightarrow \quad
\star \varphi = \begin{bmatrix}
- \bar{B} & C^\alpha & - \bar{G}
\\
- B & \bar{C}_\alpha & - G
\end{bmatrix},
\end{equation}
where, again, two successive $\star$--operations on $\varphi$
yield $\star (\star \varphi) = \pm \varphi$.

Let us now give the transformations rules for the generators of the basic
cohomology, which, just as in (\ref{1.2}), will be denoted by
$\Omega^\alpha$ (BRST operator), $^\star \Omega^\alpha =
\pm \star \Omega^\alpha \star$ (co--BRST operator) and $W^{\alpha\beta} =
\{ \Omega^\alpha, \,^\star \Omega^\beta \}$ (BRST Laplacian). Thereby,
$\Omega^\alpha$ and $^\star \Omega^\alpha$ include besides the shift and
co--shift symmetries, $Q^\alpha$ and $^\star Q^\alpha$, the
ghost--dependent ordinary gauge symmetries $\delta_G(C^\alpha)$ as well.

The transformation rules for the {\it off--shell} BRST symmetries
$\Omega^\alpha$ are
\begin{alignat}{2}
&\Omega^\alpha A_a = \psi_a^\alpha + \partial_a C^\alpha,
&\qquad
&\Omega^\alpha \psi_a^\beta = - i \epsilon_{ab} \partial^b M^{\alpha\beta},
\nonumber
\\
&\Omega^\alpha V_a = - i \psi_a^\alpha,
&\qquad
&\Omega^\alpha M_{\beta\gamma} = 2 i \delta^\alpha_{~[\beta} \zeta_{\gamma]},
\nonumber
\\
&\Omega^\alpha \zeta_\beta = i \delta^\alpha_{~\beta} B,
&\qquad
&\Omega^\alpha C^\beta = 0,
\nonumber
\\
&\Omega^\alpha \eta_\beta = i \delta^\alpha_{~\beta} G,
&\qquad
&\Omega^\alpha \bar{C}_\beta = \hbox{$\frac{1}{2}$} i
\delta^\alpha_{~\beta} (G - \bar{G}),
\nonumber
\\
&\Omega^\alpha B = 0,
&\qquad
&\Omega^\alpha \bar{B} = 0,
\nonumber
\\
&\Omega^\alpha G = 0,
&\qquad
&\Omega^\alpha \bar{G} = 0.
\end{alignat}
From combining $\Omega^\alpha$ with the Hodge--type
$\star$--symmetry (\ref{2.2}) and (\ref{3.3}) one gets the
corresponding transformation rules for the {\it off--shell}
co--BRST symmetries $^\star \Omega^\alpha$.

Then, for the non--vanishing transformations generated by the BRST Laplacian
$W^{\alpha\beta}$ one obtains
\begin{align*}
&W^{\alpha\beta} A_a = - 2 \partial_a M^{\alpha\beta},
\\
&W^{\alpha\beta} M_{\gamma\delta} = - 2 i \delta^\alpha_{~[\gamma}
\delta^\beta_{~\delta]} (G - \bar{G}),
\\
&W^{\alpha\beta} \psi_a^\gamma = 2 i \epsilon^{\alpha\beta\gamma\delta}
( \partial_a \zeta_\delta +
\epsilon_{ab} \partial^b \eta_\delta ),
\end{align*}
which, after elimination of $G$ and $\bar{G}$ through their
equations of motion, agree, as promised, with (\ref{3.2}). Let us
emphasize, that the BRST Laplacian $W^{\alpha\beta}$, in contrast
to \cite{1}, does {\it not} vanish on--shell, due to the presence of
scalar fields $M^{\alpha\beta}$. This is a consequence of the fact that
the topological nature of our model is actually encoded in the shift and
co--shift symmetries $Q^\alpha$ and $^\star Q^\alpha$, and not in the
vanishing of the BRST Laplacian as in \cite{5}.

Furthermore, by a straightforward calculation it can be verified
that $\Omega^\alpha$ and $^\star \Omega^\alpha$ actually leave the
action (\ref{3.3}) invariant, i.e., it holds $(\Omega^\alpha,
\,^\star \Omega^\alpha) S^{(N_T = 8)} = 0$. Both operators,
together with $W^{\alpha\beta}$, satisfy the following BRST
complex,
\begin{gather*}
\{ \Omega^\alpha, \Omega^\beta \} = 0,
\qquad
W^{\alpha\beta} = \{ \Omega^\alpha, \,^\star \Omega^\beta \} \neq 0,
\qquad
\{ \,^\star \Omega^\alpha, \,^\star \Omega^\beta \} = 0,
\\
[ W^{\alpha\beta}, \Omega^\gamma ] \doteq 0,
\qquad
[ W^{\alpha\beta}, \,^\star \Omega^\gamma ] \doteq 0.
\end{gather*}
Obviously, this basic cohomology is analogous to the de Rham cohomology:
The both nilpotent BRST and co--BRST operators,
$\Omega^\alpha$ and $^\star \Omega^\alpha = \pm \star \Omega^\alpha \star$,
being interrelated by the duality $\star$--operation, correspond to the
exterior and the co--exterior derivatives,
$d$ and $\delta = \pm \star d \star$, respectively, and the BRST Laplacian
$W^{\alpha\beta} = \{ \Omega^\alpha, \,^\star \Omega^\alpha \}$ is the
analogue of $\Delta = \{ d, \delta \}$, so that we have indeed a perfect
example of a Hodge type cohomological theory in two dimensions.

\bigskip
\begin{flushleft}
{\large{\bf 4. Topological observables}}
\end{flushleft}
\medskip
As already pointed out earlier, when the gauge--fixed action
(\ref{3.3}) is formulated on a compact two--dimensional Riemannian
manifold we break down the global symmetry group $SU(4)$ in such a
way that the resulting action splits into a topological term
($\Omega$--cocycle) and a $\Omega$--exact term, and that the
discrete Hodge type $\star$--symmetry is preserved, mapping
$\Omega$ to $^\star \Omega$. Alternatively, the same action can be
also obtained from the cohomological action (\ref{2.10}) by adding
the usual gauge--fixing and (anti)ghost dependent terms, see Eq.
(\ref{2.1}), and by introducing an appropriate set of auxiliary
fields.

With regard to this, let us note two unusual features of the
action (\ref{2.10}) which are relevant for the construction of
two--dimensional observables of that topological model. Its most
striking property is that the both, shift and co--shift symmetry
$Q$ and $^\star Q$, are not equivariantly nilpotent (due to the
absence of the usual ghost for ghost field $\phi$) but, on--shell,
rather they are strictly nilpotent even prior to the introduction
of the ghost and anti--ghost fields $C$ and $\bar{C}$.

Another remarkable property is that $A_a - i V_a$ and $A_a + i V_a$ are
invariant under one of the supercharges, namely $Q$ in the former and
$^\star Q$ in the latter case. Thus, for the both BRST and co--BRST
operators $\Omega$ and $^\star \Omega$ one should expect the
existence of two different sets of observables, depending either on
$A_a - i V_a$ in the former case or on $A_a + i V_a$ in the latter case.
In fact, these observables can be constructed in a similar way as
in the case of the topological sigma models \cite{14}. Therefore,
we shall omit any details and simply quote the results.

To begin with, we first associate the zero--forms $W_0$ and
$^\star W_0$ to the BRST and co--BRST transforms $\Omega \bar{C}$ and
$^\star \Omega \bar{C}$ (which, on--shell, correspond to the
gauge--fixing function $\partial^a A_a$), respectively, via the BRST and
co--BRST invariant ghost field $C$. These zero--forms can be used as
building blocks for constructing the following both sets of $k$--forms,
$W_k$ and $^\star W_k$, being interrelated by the Hodge type
$\star$--operation,
\begin{align*}
&W_0 = (\Omega \bar{C}) C,
\\
&W_1 = dx^\mu (
(\Omega \bar{C}) (A_\mu - i V_\mu) -
C \partial_\mu \bar{C} ),
\\
&W_2 = dx^\mu \wedge dx^\nu (
\bar{C} \partial_\mu (A_\nu - i V_\nu) -
(A_\mu - i V_\mu) \partial_\nu \bar{C} ),
\\
\intertext{and}
&^\star W_0 = (\,^\star \Omega \bar{C}) C,
\\
&^\star W_1 = \delta x^\mu (
(\,^\star \Omega \bar{C}) (A_\mu + i V_\mu) -
C \partial_\mu \bar{C} ),
\\
&^\star W_2 = \delta x^\mu \wedge \delta x^\nu (
\bar{C} \partial_\mu (A_\nu + i V_\nu) -
(A_\mu + i V_\mu) \partial_\nu \bar{C} ),
\end{align*}
$dx^\mu = e^\mu_{\!~~a} dx^a$ and $\delta x^\mu =
\epsilon^{\mu\nu} e_{\nu a} dx^a$, with $e^\mu_{\!~~a}$ being the
two--bein on a smooth connected, oriented Riemannian manifold $M$
endowed with metric $g^{\mu\nu}$. Here, $d = dx^\mu \partial_\mu$
and $\delta = dx^\mu \epsilon_{\mu\nu} \partial^\nu$ are the
exterior and co--exterior derivative, respectively.

These $k$--forms obey the following recursion relations, which are
typical for any topological gauge theory,
\begin{alignat}{4}
0 &= \Omega W_0,
&\qquad
d W_0 &= \Omega W_1,
&\qquad
d W_1 &= \Omega W_2,
&\qquad
d W_2 &= 0,
\nonumber
\\
0 &= \,^\star \Omega \,^\star W_0,
&\qquad
\delta \,^\star W_0 &= \,^\star \Omega \,^\star W_1,
&\qquad
\delta \,^\star W_1 &= \,^\star \Omega \,^\star W_2,
&\qquad
\delta \,^\star W_2 &= 0.
\label{4.1}
\end{alignat}
Now, if $\gamma$ is a $k$--dimensional homology cycle, $\partial \gamma = 0$,
on $M$ then the integrated $k$--forms
\begin{equation*}
I_k(\gamma) = \int_{\gamma} W_k,
\qquad
\end{equation*}
by virtue of (\ref{4.1}), are $\Omega$--invariant,
\begin{equation*}
\Omega I_k(\gamma) = \int_\gamma \Omega W_k = \int_\gamma d W_{k - 1} = 0,
\qquad
k > 0.
\end{equation*}
Moreover, if $\beta = \partial \alpha$ is the boundary of a $(k +
1)$--dimensional surface, $k < 2$, so that $\beta$ is trivial in
homology, then $I_k(\gamma)$ depends only upon the homology class
of $\gamma$ up to a $\Omega$--exact term,
\begin{equation*}
I_k(\gamma + \partial \alpha) = \int_{\gamma + \partial \alpha} W_k =
I_k(\gamma) + \int_\alpha d W_k = I_k(\gamma) +
\int_\alpha \Omega W_{k + 1} = I_k(\gamma).
\end{equation*}
Finally, following \cite{14}, one can introduce gauge invariant correlation
functions of arbitrary products of the $I_k(\gamma)$,
\begin{equation*}
Z(\gamma_1, \ldots, \gamma_r) = \int D \varphi\, {\rm exp}( - S(\varphi) )
\prod_{i = 1}^r \int_{\gamma_i} W_{k_i}(\varphi),
\end{equation*}
which, by construction, both are $\Omega$--invariant and invariant
under metric deformations which preserve the holonomy structure.
The same constructions hold for the $k$--forms $^\star W_k$.

Summarizing, we have shown that, on--shell, the vanishing of the
BRST Laplacian of the Hodge theory proposed in \cite{5} can be avoided,
if we view the Maxwell action as the classical part of a more involved
cohomological action, which is obtained by a $N_T = 8$ topological twist
of $N = 16$, $D = 2$ super Maxwell theory with global symmetry group $SU(4)$.
Then, the complete gauge--fixed cohomological action is of Witten type and
the underlying basic cohomology is really of Hodge type.
The non--Abelian case will be presented elsewhere.
\bigskip

\noindent {\Large \bf {Acknowledgment}} The authors grateful
acknowledge discussions with M. Blau which led them to the
foregoing considerations.



\bigskip
\begin{thebibliography}{100}
\bibitem{1} A. Migdal,
            {\it Zh. Eksp. Theo. Fiz.} {\bf 69} (1975) 810
            [Sov. Phys. JETP. {\bf 42} (1975) 743]
\bibitem{2} D. Gross,
            {\it Nucl. Phys.} {\bf B 400} (1993) 161;
            D. Gross and W. Taylor IV,
            {\it Nucl. Phys.} {\bf B 400} (1993) 181;
            {\it Nucl. Phys.} {\bf B 403} (1993) 395
\bibitem{3} E. Witten,
            {\it J. Geom. Phys.} {\bf 9} (1992) 303
\bibitem{4} E. Witten,
            {\it Commun. Math. Phys.} {\bf 141} (1991) 153
\bibitem{5} R. P. Malik,
            {\it J. Phys. A: Math. Gen.} {\bf 33} (2000) 2437;
            {\it Int. J. Mod. Phys.} {\bf A 15} (2000) 1685;
            {\it Mod. Phys. Lett.} {\bf A 14} (1999) 1937;
            {\it Mod. Phys. Lett.} {\bf A 15} (2000) 2079;
            {\it Mod. Phys. Lett.} {\bf A 16} (2001) 477;
            {\it J. Phys. A: Math. Gen.} {\bf 34} (2001) 4167
\bibitem{6} D. Birmingham, M. Blau, M. Rakowski and G. Thompson,
            {\it Topological Field Theories}, Phys. Repts. {\bf 209} (1991) 129
\bibitem{7} K. Nishijima,
            {\it Prog. Theor. Phys.} {\bf 80} (1988) 897
\bibitem{8} A. V. Razumov and G. N. Rybkin,
            {\it Nucl. Phys.} {\bf B 332} (1990) 209
\bibitem{9} J. W. van Holten,
            {\it Aspects of BRST quantization}, Lectures at Summer School
            {\it Geometry and Topology in Physics} (Rot a.d. Rot,
            Germany), hep-th/0201124, and references therein
\bibitem{10} S. Ouvry, R. Stora and P. van Baal,
            {\it Phys. Lett.} {\bf B 229} (1989) 159
\bibitem{11} B. Geyer and D. M\"ulsch,
             {\it Int. J. Mod. Phys.} {\bf A 17} (2002) 1183;
             {\it Nucl. Phys.} {\bf B 662} (2003) 531
\bibitem{12} M. Blau and G. Thompson,
             {\it Nucl. Phys.} {\bf B 492} (1997) 545
\bibitem{13} B. Geyer and D. M\"ulsch,
             {\it Phys. Lett.} {\bf B 518} (2001) 181
\bibitem{14} E. Witten,
             {\it Commun. Math. Phys.} {\bf 118} (1988) 411
\end{thebibliography}
\end{document}